# A new exceptional point condition for coupled microresonators with coupled mode theory in space

Kunpeng Zhu,[1] Xiaoyan Zhou,[1,*] Yinxin Zhang,[1] Zhanhua Huang,[1] and Lin Zhang[1,2,*]

[1]State Key Laboratory of Precision Measuring Technology and Instruments, Key Laboratory of Opto-electronic Information Technology of Ministry of Education, Tianjin Key Laboratory of Integrated Opto-electronics Technologies and Devices, School of Precision Instruments and Opto-electronics Engineering, Tianjin University, Tianjin 300072, China
[2]Peng Cheng Laboratory, Shenzhen 518038, China
*Corresponding authors: xiaoyan_zhou@tju.edu.cn, lin_zhang@tju.edu.cn

**We derive new exceptional point (EP) conditions of the coupled microring resonators using coupled mode theory in space, a more accurate approach than the commonly used coupled mode theory in time. Transmission spectra around EPs obtained from the two models have been compared on two material platforms, revealing non-negligible deviations. Our analysis provides a guide for accurately determining parameter sets of coupled microrings at EPs and deepens our understanding on parity-time-symmetric coupled resonators at EPs.**

In quantum mechanics, non-Hermitian Hamiltonians generally exhibit complex eigenvalues [1]. The discovery that non-Hermitian systems can also possess real spectra under the condition of parity-time (PT) symmetry has drawn considerable research attention [2–4]. Exceptional points (EPs), also known as non-Hermitian degeneracies, are singularities where two or more eigenvalues and associated eigenstates simultaneously coalesce [5,6]. Due to the analogy between the Schrödinger equation and the paraxial wave equation, optical and photonic systems can be engineered to operate at EPs under the PT symmetry condition, i.e., the real part of the complex refractive index distribution is an even function of the position, while the imaginary part, representing gain or loss, is an odd function [7,8]. Coupled microresonators offer an excellent platform for the realization of EPs, when material gain is introduced to one resonator [9–15]. Thanks to the high-quality-factor resonance, microresonators working at EPs have been leveraged to achieve stable single-longitudinal-mode lasing [9,12,13], unidirectional transmission [10,11], ultra-sensitive sensing [14–20], and coherent perfect absorption [21].

Although theoretical predictions indicate significantly enhanced device performance [14], experimental results have sometimes fallen short of these predictions, achieving an enhancement factor of 4 [16] and 20 [17] in gyroscope, 2.5 in single-particle detection [18], and 23 in temperature measurement [15]. The discrepancies between theoretical predictions and experimental results can be attributed to fabrication uncertainties of the devices. In addition, gain saturation [22] and fluctuation noise [23] have been discussed as potential reasons for the discrepancies. In fact, before taking these practical factors into account, it is critical to examine the theoretical accuracy of the design tool for coupled-microresonator-based EP devices. Typically, coupled mode theory in time (CMTT) [24] is employed to model the coupled microresonators and to determine the EP conditions, owning to its similarity to the Schrödinger equation. In contrast, coupled mode theory in space (CMTS, also called the transfer matrix method) [25,26] is known to be more accurate than the former, particularly when coupling and loss coefficients as well as resonator radii are large [27]. However, the CMTS has seldomly been used to explore the EP conditions of the coupled microresonators.

In this Letter, we derive a new set of the EP conditions of the coupled ring resonators using CMTS, a more accurate approach than commonly used CMTT. The in-theory discrepancies between the CMTS and the CMTT are identified by examining the transmission characteristics of the coupled microrings, considering both low and high loss/gain material platforms. Intriguingly, our findings reveal that, with CMTS, the striking spectral feature of EP devices, including the infinite resonant transmission and the zero bandwidth, is absent with parameter sets derived from CMTT. This non-negligible deviation between the two approaches highlights the critical role of precisely modeling the coupled microresonators in accurately determining the EP conditions and potentially reducing the inconsistencies between theory and experiments.

We start with a brief review of the CMTT for coupled microring resonators [24], as shown in Fig. 1(a). The time evolution of the fields in the two resonators is described as follows:

$$\begin{cases} \dfrac{da_1}{dt} = -i\omega_1 a_1 - g_1 a_1 + i\mu a_2 + i\sqrt{\gamma_{c1}} E_{in} \\ \dfrac{da_2}{dt} = -i\omega_2 a_2 - g_2 a_2 + i\mu a_1 \end{cases}, \quad (1)$$

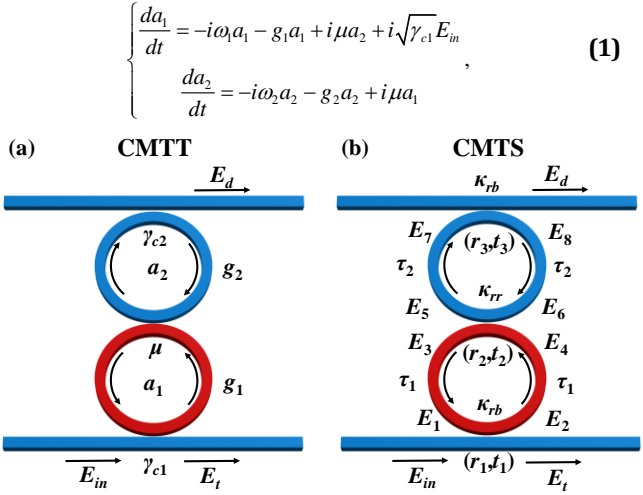

**Fig. 1.** Modeling PT-symmetric coupled microring resonators using (a) CMTT and (b) CMTS. The ring with gain is highlighted in red.

where $E_{in}$ is the electric field at the input waveguide, $a_j$, $\omega_j$, and $g_j$ are the energy amplitude, resonant angular frequency, and gain/loss rate in the ring with gain ($j = 1$, red colored) or the ring with loss ($j = 2$, blue colored), $g_{1,2}$ are related to the intrinsic gain/loss rates in the rings, $\gamma_{1,2}$, and waveguide-ring coupling rates, $\gamma_{c1,c2}$, by $g_{1,2} = \gamma_{1,2}/2 + \gamma_{c1,c2}/2$, and $\mu$ is the coupling strength between the rings. For simplicity, the two microrings are assumed to have the same geometric parameters ($R_1 = R_2 = R$, $\omega_1 = \omega_2 = \omega_0$) and the same waveguide-ring coupling rate ($\gamma_{c1} = \gamma_{c2} = \gamma_c$). In this regard, $\gamma_{1,2}$ are related to the gain/loss coefficients in the waveguides consisting of the microrings, $\alpha_{1,2}$, by $\alpha_{1,2} 2\pi R = \gamma_{1,2}\tau_R$, where $\tau_R$ is the round-trip time of each ring. $\gamma_c$ and $\mu$ are related to the ring-bus waveguide power coupling coefficient, $\kappa_{rb}$, and ring-ring power coupling coefficient, $\kappa_{rr}$, by $\gamma_c = \kappa_{rb} v_g/(2\pi R)$ and $\mu = \sqrt{\kappa_{rr}} v_g/(2\pi R)$, respectively, where $v_g$ is the group velocity of the guided mode.

The transfer function can be derived from Eq. (1) as

$$\frac{E_t}{E_{in}} = \frac{[i(\omega-\omega_1+\frac{\gamma_c}{2}-\frac{\gamma_1}{2})][i(\omega-\omega_2)-g_2]+\mu^2}{[i(\omega-\omega_1)-g_1][i(\omega-\omega_2)-g_2]+\mu^2}. \quad (2)$$

By setting the denominator to zero, the eigenfrequencies of the coupled resonators are obtained as [28]

$$\omega_\pm - \omega_0 = -\frac{i(g_1+g_2)}{2} \pm \sqrt{\mu^2 - \left(\frac{g_1-g_2}{2}\right)^2}. \quad (3)$$

The system exhibits PT symmetry when $g_1 + g_2 = 0$, and it works at EPs when $\mu^2 = (g_1-g_2)^2/4$. According to these relationships, at EPs, the power coupling coefficients ($\kappa_{rr0}$ and $\kappa_{rb0}$) and the gain/loss coefficients should satisfy

$$\begin{cases} \kappa_{rr0} = \left(\dfrac{\alpha_1 L/2 - \alpha_2 L/2}{2}\right)^2 \\ \kappa_{rb0} = -\alpha_1 L/2 - \alpha_2 L/2 \end{cases}, \quad (4)$$

where $L$ is the circumference of the rings. From Eq. (4), we note that, for coupled microrings with fixed geometries and gain/loss coefficients, $\kappa_{rr}$ and $\kappa_{rb}$ can be tailored to configure the system to work at EPs. This can be practically implemented by adjusting the coupling distances [10,11].

Alternatively, the coupled microring resonators can be modeled using the CMTS [25,26], as shown in Fig. 1(b). The fields at various locations in the rings and waveguides, $E_{1-11}$, satisfy

$$\begin{bmatrix} E_2 \\ E_t \end{bmatrix} = \begin{bmatrix} r_1 & it_1 \\ it_1 & r_1 \end{bmatrix}\begin{bmatrix} E_1 \\ E_{in} \end{bmatrix}; \begin{bmatrix} E_5 \\ E_3 \end{bmatrix} = \begin{bmatrix} r_2 & it_2 \\ it_2 & r_2 \end{bmatrix}\begin{bmatrix} E_6 \\ E_4 \end{bmatrix}; \begin{bmatrix} E_d \\ E_8 \end{bmatrix} = \begin{bmatrix} r_3 & it_3 \\ it_3 & r_3 \end{bmatrix}\begin{bmatrix} 0 \\ E_7 \end{bmatrix}$$

$$\begin{bmatrix} E_4 \\ E_1 \\ E_7 \\ E_6 \end{bmatrix} = \begin{bmatrix} \tau_1 e^{i\varphi_1} & 0 & 0 & 0 \\ 0 & \tau_1 e^{i\varphi_1} & 0 & 0 \\ 0 & 0 & \tau_2 e^{i\varphi_2} & 0 \\ 0 & 0 & 0 & \tau_2 e^{i\varphi_2} \end{bmatrix}\begin{bmatrix} E_2 \\ E_3 \\ E_5 \\ E_8 \end{bmatrix}, \quad (5)$$

where $r_{1,2,3}$ and $t_{1,2,3}$ are the amplitude transmission and coupling coefficients at the coupling regions shown in Fig. 1(b), respectively, and $r_{1,2,3}^2 + t_{1,2,3}^2 = 1$, $\tau_{1,2}$ and $\varphi_{1,2}$ are the amplitude transmissions and phase shifts in half of the rings, and $\varphi_{1,2} = \pi n_{eff} L/\lambda_0$ ($\lambda_0$ is the resonant wavelength and $n_{eff}$ is the effective index of the guided mode).

From Eq. (5), the transfer function is derived to be

$$\frac{E_t}{E_{in}} = \frac{r_3 e_1^2 e_2^2 \tau_1^2 \tau_2^2 - r_2 e_1^2 \tau_1^2 - r_1 r_2 r_3 e_2^2 \tau_2^2 + r_1}{r_1 r_3 e_1^2 e_2^2 \tau_1^2 \tau_2^2 - r_1 r_2 e_1^2 \tau_1^2 - r_2 r_3 e_2^2 \tau_2^2 + 1}, \quad (6)$$

where $e_{1,2} = e^{i\varphi_{1,2}}$. The EP condition in the CMTS formalism is not immediately apparent. We present the new EP conditions below

$$\begin{cases} r_1 r_3 \tau_1^2 \tau_2^2 - r_1 r_2 \tau_1^2 - r_2 r_3 \tau_2^2 + 1 = 0 \\ \tau_1^2 r_1 = 1/(\tau_2^2 r_3) \end{cases}. \quad (7)$$

Similar to the derivation of the eigenfrequencies in CMTT, the first equation of Eq. (7) is obtained by setting the denominator of Eq. (6) to zero. The second equation of Eq. (7) is derived from the PT symmetry condition, i.e., the round-trip amplitude transmission in the first ring must be equal to the reciprocal of that in the second ring. In fact, it is the same as $g_1 + g_2 = 0$ in CMTT when $\tau_{1,2}$ and $r_{1,3}$ approach 1 (see Supplement 1). Using $r_1 = r_3$ and substituting the second equation from Eq. (7) into the first, we obtain

$$2 - r_2 \tau_1/\tau_2 - r_2 \tau_2/\tau_1 = 0. \quad (8)$$

By combining Eqs. (7) and (8), we arrive at

$$\begin{cases} r_1 = (\tau_1 \tau_2)^{-1} \\ r_2 = 2\tau_1 \tau_2/(\tau_1^2 + \tau_2^2) \end{cases}. \quad (9)$$

According to $\kappa_{rb} + r_1^2 = 1$, $\kappa_{rr} + r_2^2 = 1$, and $\tau_{1,2} = \exp(-\alpha_{1,2}L/4)$, the relations between the power coupling coefficients ($\kappa'_{rr0}$ and $\kappa'_{rb0}$) and the gain/loss coefficients at EPs, within the framework of CMTS, are derived as

$$\begin{cases} \kappa'_{rr0} = \dfrac{(\exp(-\alpha_1 L/2) - \exp(-\alpha_2 L/2))^2}{(\exp(-\alpha_1 L/2) + \exp(-\alpha_2 L/2))^2} \\ \kappa'_{rb0} = \dfrac{\exp(-\alpha_1 L/2 - \alpha_2 L/2) - 1}{\exp(-\alpha_1 L/2 - \alpha_2 L/2)} \end{cases}. \quad (10)$$

In fact, Eq. (10) can be simplified to Eq. (4) by taking the first-order Taylor series approximation at $\alpha_{1,2}L = 0$ (see Supplement 1). This means that the EP conditions derived from CMTT is less accurate, when $\alpha_{1,2}L$ is much larger than 0.

To quantitatively compare the differences of the two models, we obtain the power transmission spectra at EPs using Eq. (4). We consider two distinct material platforms with different gain (as well as loss) features: the SiN platform and the III-V semiconductor platform. On the SiN platform, optical gain can be achieved by doping the resonators with $Er^{3+}$ ions [10], which is demonstrated in a range of 0.5-20 dB/cm [29,30]. We assume that the first ring has a gain coefficient of 10 dB/cm, and the second one has a loss coefficient of -3 dB/cm [30]. The radii of both resonators are set to 200 µm. According to Eq. (4), power coupling coefficient $\kappa_{rr0}$ and $\kappa_{rb0}$ are calculated to be 0.0088 and 0.1013, respectively. On the III-V platform, a high material gain coefficient of ~1000 dB/cm can be attained in multiple quantum wells [12,13,31], and a loss coefficient of ~-200 dB/cm can be introduced by depositing lossy materials such as Ge/Cr on top of the waveguides [12]. Here, we use a gain coefficient of 200 dB/cm and a loss coefficient of -50 dB/cm. The radii are set to 20 μm. The power coupling coefficients $\kappa_{rr0}$ and $\kappa_{rb0}$ obtained from Eq. (4) are 0.0327 and 0.2169, respectively.

In Fig. 2, with CMTT, the real and imaginary parts of the eigenfrequencies, Re($\omega$-$\omega_0$) and Im($\omega$-$\omega_0$), are presented as a function of $\kappa_{rr}$ for the two platforms near the EPs. We note that, with CMTS, it is difficult to analytically derive the eigenfrequencies. Thus, numerically, we obtain the frequency splitting $\Delta\omega$ and the full-width-half-maximum (FWHM) in the transmission spectrum from CMTS. Since Im($\omega$-$\omega_0$) is equal to half of the FWHM in the single microring. By analogy, FWHM/2 and -FWHM/2 are plotted in the Figs. 2(a) and 2(b) as counterparts of Im($\omega$-$\omega_0$).

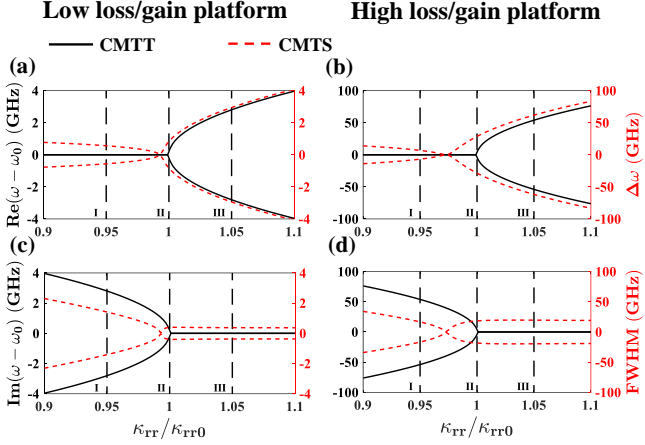

**Fig. 2.** (a) and (b) Re($\omega$-$\omega_0$) in CMTT and $\Delta\omega$ in CMTS around the EP conditions derived from CMTT, i.e., $\kappa_{rr} = \kappa_{rr0}$. (c) and (d) Im($\omega$-$\omega_0$) in CMTT and FWHM in CMTS around $\kappa_{rr} = \kappa_{rr0}$.

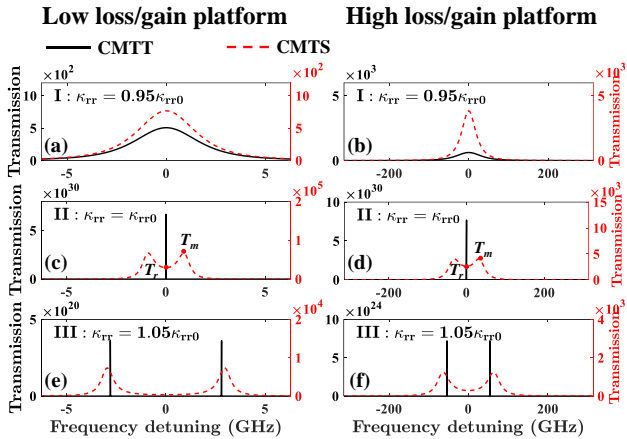

**Fig. 3.** Power transmission spectra of two approaches using parameter sets calculated from CMTT.

For the low loss/gain platform, Figs. 2(a) and 2(c) show that, the Re($\omega$-$\omega_0$) coalesces for $\kappa_{rr} < \kappa_{rr0}$ and bifurcates for $\kappa_{rr} > \kappa_{rr0}$, while the Im($\omega$-$\omega_0$) exhibits inverse behavior, corresponding to the PT-broken regime (I) and the PT-symmetric regime (III), respectively. At $\kappa_{rr} = \kappa_{rr0}$, both Re($\omega$-$\omega_0$) and Im($\omega$-$\omega_0$) coalesce, and the coupled microrings are operated at the EP (II). Note that these EP conditions are derived from CMTT. When the same $\kappa_{rr}$ is employed in CMTS ($\kappa'_{rr} = \kappa_{rr0}$), both $\Delta\omega$ and FWHM exhibit bifurcation instead of coalescence. Furthermore, neither $\Delta\omega$ nor FWHM converges to zero for $\kappa'_{rr} \neq \kappa'_{rr0}$, in contrast to Re($\omega$-$\omega_0$) = 0 for $\kappa_{rr} < \kappa_{rr0}$ and Im($\omega$-$\omega_0$) = 0 for $\kappa_{rr} > \kappa_{rr0}$ with CMTT. In Figs. 2(b) and 2(d), similar phenomena are observed for the high loss/gain platform, albeit with larger $\Delta\omega$ and FWHM values. These observations in Fig. 2 indicate a great distinction between EPs from CMTS and those from CMTT.

Figure 3 shows the power transmission spectra ($|E_t|^2/|E_{in}|^2$) obtained using the two models at three different $\kappa_{rr}$ values, corresponding to regimes (I), (II), and (III) in Fig. 2, respectively. When $\kappa_{rr} = 0.95\kappa_{rr0}$, as shown in Fig. 3(a), the PT-symmetric phase is observed in both approaches, characterized by a single finite peak with a nonzero FWHM. In Fig. 3(c), when $\kappa_{rr} = \kappa_{rr0}$, the transmission at zero frequency detuning as obtained from CMTT exhibits infinite transmission and zero FWHM, characteristic of the EPs. In contrast, the peak transmission obtained from CMTS is significantly lower, with noticeable $\Delta\omega$ and nonzero FWHM. As shown in Fig. 3(e), when $\kappa_{rr} = 1.05\kappa_{rr0}$, a transmission spectrum from CMTT is with nonzero $\Delta\omega$ and zero FWHM, corresponding to the broken-PT phase; while, with CMTS, the transmission spectrum exhibits slightly different $\Delta\omega$ and nonzero FWHM. Similarly, on the high loss/gain platform as shown in Figs. 2(b), 2(d), and 2(f), the PT-symmetric phase, EP, and broken-PT phase are observed in CMTT; whereas EP and PT-symmetric phase are absent in CMTS, and this indicates discrepancies in EPs derived from the two approaches.

From the above, it is noted that around the EP conditions set by CMTT, one cannot produce the EP effect by simply varying $\kappa_{rr}$ with CMTS. Thus, we perform a scan of both $\kappa_{rb}$ and $\kappa_{rr}$ near the EPs from

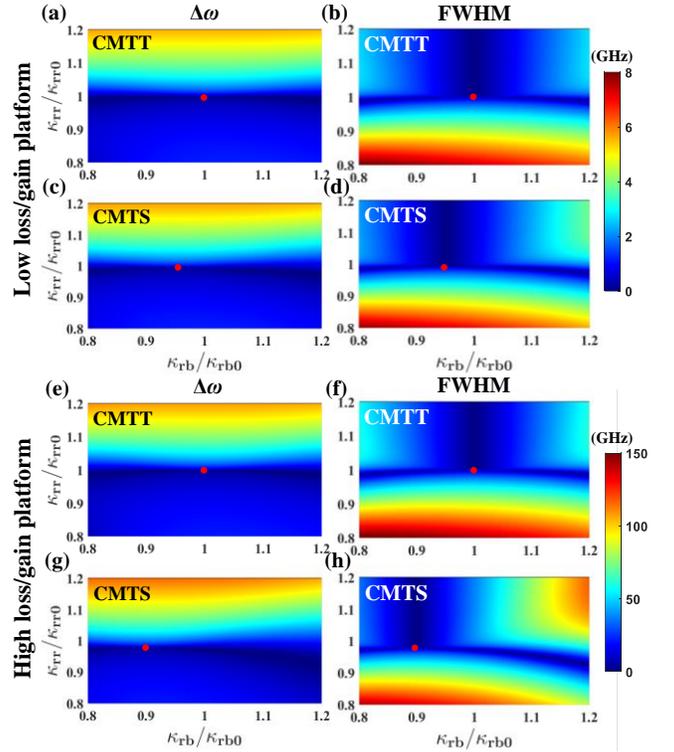

**Fig. 4.** $\Delta\omega$ and FWHM for different $\kappa_{rr}$ and $\kappa_{rb}$ near EPs. $\kappa_{rb0}$ and $\kappa_{rr0}$ are power coupling coefficients at EPs calculated from CMTT.

CMTT and plot $\Delta\omega$ and FWHM in Fig. 4. The EP is denoted by a red dot in each plot. On the low loss/gain platform, as shown in Figs. 4(a) and 4(b), the $\Delta\omega$ and FWHM are zero when $\kappa_{rb} = \kappa_{rb0}$ and $\kappa_{rr} = \kappa_{rr0}$ in CMTT. In contrast, as shown in Figs. 4(c) and 4(d), the EP features are observed at a smaller $\kappa_{rb}$ value, i.e., $\kappa'_{rb0} < \kappa_{rb0}$. According to Eq. (10), $\kappa'_{rb0} = 0.0963 \approx 0.95\kappa_{rb0}$, and $\kappa'_{rr0} = 0.0088 \approx \kappa_{rr0}$. The EP locations in the 2D plots match these calculated results well. On the high loss/gain platform, similar shifts in EP obtained from CMTS are observed, which again validates the EP conditions given in Eq. (10), i.e., $\kappa'_{rb0} = 0.1949 \approx 0.90\kappa_{rb0}$, and $\kappa'_{rr0} = 0.0319 \approx 0.98\kappa_{rr0}$. From Fig. 4, it is noted that the $\kappa'_{rb0}$ and $\kappa'_{rr0}$ at EPs from CMTS deviate from those from CMTT, and the differences are more visible for high loss/gain platform. In fact, many experimental parameter sets fall within the relatively inaccurate parameter spaces of the CMTT. For instance, the devices in Refs. [12] and [15] exhibit large gains of 1164 dB/cm and 434 dB/cm and losses of -1164 dB/cm and -434 dB/cm. The parameters in Ref. [15] correspond to $\kappa'_{rr0} = 0.0937 \approx 0.94\kappa_{rr0}$, and $\kappa'_{rb0} = \kappa_{rr0} = 0$. The discrepancies between two approaches could partially account for the non-ideal experimental observations of EP-based devices.

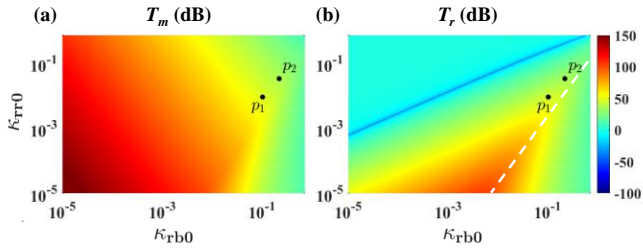

**Fig. 5.** Transmission maps calculated from CMTS at different $\kappa_{rb0}$ and $\kappa_{rr0}$ that satisfy the EP conditions given by CMTT. (a) Maximum power transmission $T_m$. (b) Resonant power transmission $T_r$.

Figure 5 gives an overview of the transmission differences between the two approaches. We obtain the transmission spectra with CMTS, using a varied parameter set of $\kappa_{rb0}$, $\kappa_{rr0}$, and gain/loss coefficients that satisfy the EP conditions by CMTT (the gain/loss plots are given in Supplement 1). Both maximum power transmission, $T_m$, and power transmission at zero frequency detuning, $T_r$, are shown in Figs. 5(a) and 5(b), respectively, as they can be different due to frequency splitting. The parameter sets of the low loss/gain platform and the high loss/gain platform in the above are labelled as $p_1$ and $p_2$, respectively. In Fig. 5(a), $T_m$ are small in case of large $\kappa_{rb0}$ and $\kappa_{rr0}$, and approach infinity (an EP feature) as $\kappa_{rb0}$ and $\kappa_{rr0}$ decrease to zero (corresponding to the very low loss/gain case). This means that CMTS reduces to CMTT when the coupling coefficients are small. On the other hand, $T_r$ in Fig 5(b) shows significant deviations between the two models. Only the coupling coefficient sets in the region below the white dashed line result in a single transmission peak around the resonant frequency, and the other combinations of $\kappa_{rb0}$ and $\kappa_{rr0}$ give mode-splitting transmission lineshapes, which are similar to those in Figs. 3(c) and 3(d), leading to a smaller $T_r$ than $T_m$. $T_r$ can even be zero for certain sets of $\kappa_{rb0}$ and $\kappa_{rr0}$, as shown by a blue line in Fig. 5(b).

Although CMTT is computationally fast and accurate in most situations in contrast to finite difference time domain, it will be more realistic to consider coupling loss (see Supplement 1). After the derivation in supplement, we find the $\kappa'_{rb0}$ is unchanged and $\kappa'_{rr0}$ becomes larger.

In conclusion, we derive new EP conditions for coupled ring resonators using CMTS. The pronounced discrepancies between two models in terms of transmission spectra reveal the importance of carefully examining the theoretical accuracy of EP models, especially with large coupling, gain and loss coefficients as well as resonator radii. We find that CMTS is more accurate for determining the EP conditions, potentially providing a venue to approach closer to the EPs in practical applications.

**Disclosures.** The authors declare no conflicts of interest.

**Data availability.** Data underlying the results presented in this paper are not publicly available at this time but may be obtained from the authors upon reasonable request.

**Supplemental document.** See Supplement 1 for supporting content.